\documentclass{article}

\usepackage{PRIMEarxiv}

\usepackage[utf8]{inputenc} 
\usepackage[T1]{fontenc}    
\usepackage[hidelinks]{hyperref}       
\usepackage{url}            
\usepackage{booktabs}       
\usepackage{amsfonts}       
\usepackage{nicefrac}       
\usepackage{microtype}      
\usepackage{lipsum}
\usepackage{fancyhdr}       
\usepackage{graphicx}       
\usepackage{multirow}
\usepackage[numbers,sort&compress]{natbib}
\graphicspath{{media/}}     

\title{A Systematic Approach for Assessing Large Language Model’s Test Case Generation Capability
}

\author{
  Hung-Fu Chang \\
  R. B. Annis School of Engineering \\
  University of Indianapolis \\
  Indianapolis\\
  \texttt{hchang@uindy.edu} \\
   \And
  Mohammad Shokrolah Shirazi  \\
  E. S. Witchger School of Engineering \\
  Marian University \\
  Indianapolis\\
  \texttt{mshokrolahshirazi@marian.edu} \\
}

\begin{document}
\maketitle

\begin{abstract}
Software testing ensures the quality and reliability of software products, but manual test case creation is labor-intensive. With the rise of large language models (LLMs), there is growing interest in unit test creation with LLMs. However, effective assessment of LLM-generated test cases is limited by the lack of standardized benchmarks that comprehensively cover diverse programming scenarios. To address the assessment of LLM’s test case generation ability and lacking dataset for evaluation, we propose the Generated Benchmark from Control-Flow Structure and Variable Usage Composition (GBCV) approach, which systematically generates programs used for evaluating LLMs' test generation capabilities. By leveraging basic control-flow structures and variable usage, GBCV provides a flexible framework to create a spectrum of programs ranging from simple to complex. Because GPT-4o and GPT-3-Turbo are publicly accessible models, to present real-world regular user’s use case, we use GBCV to assess LLM performance on them. Our findings indicate that GPT-4o performs better on complex program structures, while all models effectively detect boundary values in simple conditions but face challenges with arithmetic computations. This study highlights the strengths and limitations of LLMs in test generation, provides a benchmark framework, and suggests directions for future improvement.
\end{abstract}

\keywords{Software Testing; Test Case Generation; Large Language Model; GPT}

\section{Introduction}
Software testing is an important component of the software development lifecycle because the practice ensures the quality and functionality of software products \cite{tassey_2002_}. It is among the most resource-intensive aspects of software development, playing a critical role in validating that software systems meet established requirements and operate without faults \cite{grano_2018_an, tufano_2020_unit}. As a result, developers employ various rigorous testing to identify bugs to significantly reduces the risk of costly defects in production environments \cite{grano_2018_an, tufano_2020_unit, winkler_2024_investigating}. Among tests like integration test, acceptance test, regression test etc. in the software development, unit test can be regard as the most essential and widely used practices, where test cases are written to validate specific methods or components of the source code in isolation. This focused approach facilitates the early identification of defects at their source, minimizing their potential impact on the broader software system \cite{chen_2023_chatunitest, siddiq_2024_using}.

Since manual unit test case creation is a time-consuming process, consuming over 15\% of a developer’s time on average \cite{daka_2014_a}, automating test case generation from codes or requirements has become an area motivating extensive research. Various generation approaches have been suggested, as outlined in review and survey studies \cite{anand_2013_an, kaur_2012_systematic}. Practical application of automatic test case generation was also explored in past research \cite{brunetto_2021_on}. Due to the emergence of Large Language Models (LLMs), many scholars and industrial practitioners feel the significant potential of LLMs in automating software development tasks. Their ability to learn from extensive datasets and produce contextual codes to form relevant function or programs. Therefore, traditional automating coding-related tasks, such as code generation, test case creation, test execution, and code recommendation, transforms their applications or studies into possible integration with LLMs \cite{li_2024_structured, li_2023_skcoder, li_2023_acecoder, dong_2023_selfcollaboration}. That is, with their remarkable performance in generation and inference tasks, researchers can explore new applications of LLMs to software testing. In particular, unit test case generation has emerged as a promising area of research. 

Generating unit test cases with LLM introduces unique challenges. To generate effective and accurate test cases, the training and evaluation requires diverse inputs and comprehensive coverage so that the generation tests can validate functional correctness to ensure robust software components. Overcoming this challenge heavily depends on the availability of a benchmark dataset, which serves as a mean for evaluating task performance and can also be utilized for further training. An analogy can be drawn from text-to-SQL research, where datasets like Spider \cite{chen_2018_spider} and BIRD \cite{li_2023_can} are commonly used both to assess the performance of generating SQL queries from texts and to support text-to-SQL training. While previous research, such as TESTEVAL \cite{wang_2024_testeval}, proposed a benchmark by collecting Python programs from LeetCode to evaluate test case generation with LLMs, there is still a lack of a widely accepted, generic, and extendable benchmark dataset for assessing the quality, coverage, and effectiveness of tests generated by LLMs, as well as for training test case generation. This critical absence of a benchmarking dataset for LLM-based test generation limits the ability to reliably and objectively assess the performance and effectiveness of LLMs in generating accurate and functional test cases, especially when the LLM has been updated through new or different training or tuning methods. Hence, to remedy this limitation requires the development of a benchmark dataset for systematic evaluation of LLM’s test case generation ability.

Instead of directly creating a benchmarking dataset, which typically involves collecting data from practices or online sources as done for other domains, we propose an approach called Generated Benchmark from Control-Flow Structure and Variable Usage Composition (GBCV), which is designed to target any programming language for assessing test case generation. To maintain its flexibility, GBCV enables users to develop various control-flow structures and define desired variable usages, which are then used to generate programs for testing. These programs are subsequently incorporated into prompts to guide the LLM in creating unit test cases. The test cases are then applied to the programs for assessing the LLM’s test case capability. In addition, GBCV suggests generating code by starting with basic control-flow structures, which enables the incremental construction of more complex structures without being restricted to existing code data-bases. It also provides users with a way to create a comprehensive spectrum of programs, ranging from simple and small programs to large and complex ones.

This paper marks an important first step toward our ambitious goal of advancing LLM’s capabilities, through continuous improvements, in generating high-quality test cases. In this study, we focus on our scope to introducing the GBCV method, illustrating its use, and evaluating existing LLMs with this approach. Through these assessments, we not only highlight the utility of GBCV but also uncover key insights that pave the way for further research and refinement. Ultimately, our work lays the groundwork for developing more effective and reliable LLMs capable of producing robust test cases. The remainder of this paper is structured as follows: section 2 discusses related work, while section 3 details the GBCV approach and the evaluation metrics. Section 4 shows and analyzes the investigation results. Finally, section 5 concludes the study and outlines the directions for future research.

\section{Related Work}
Our research aligns with two closely related areas: the utilization of Large Language Models (LLMs) or natural language processing (NLP) techniques for automated test case generation and the development of benchmark datasets to evaluate LLM performance in generating computing artifacts. 

\subsection{Test Cases Generation via LLMs}
Test cases must be generated from certain software artifacts, which specified the inputs or expected outputs of the test cases \cite{anand_2013_an}. Traditionally, according to the types of artifacts used, test case generation techniques fall into several categories, such as symbolic execution \cite{anand_2011_heap}, search-based and evolutionary approaches \cite{fraser_2011_evolutionary, fraser_2011_evosuite}. These approaches rely on static or dynamic analysis techniques to explore a program’s control and data flow paths. These typically result in test cases that are less readable and understandable than manually written test \cite{mmoeinalmasi_2017_an} and may either lack assertions or include only generic ones \cite{panichella_2020_replication}. With the advancement of LLMs, researchers have explored their potential to automate certain software artifacts like codes, which has sparked interest in integrating LLMs into software testing, particularly for test case generation. The growing attention on LLM-based test case generation aims to improve efficiency, enable self-learning in test creation, or produce more natural-looking tests \cite{schfer_2023_an}. 

One way to use LLMs in test case generation is by emphasizing prompt creation. Yuan et al. [12] proposed a framework, ChatTester, which iteratively refines test cases by asking ChatGPT to generate follow-up prompts. This method resulted in a 34.3\% increase in compilable tests and an 18.7\% improvement in tests with correct assertions, demonstrating the potential of LLMs in enhancing test generation. Schäfer et al. \cite{schfer_2023_an} introduced TestPilot, a tool automates unit test generation using a Javascript framework without additional trainings. TestPilot leveraged OpenAI's Codex to generate unit tests by con-structing prompts with analyzing function's signature, usage snippets, and documentation, then refining prompts to improve generation outcomes. Similarly, ChatUniTest, developed by Xie et al. \cite{xie_2023_chatunitest} , extracted raw information from JAVA codes, converted into an Abstract Syntax Tree (AST). It utilizes static analysis during the preparation stage and applies an adaptive focal context mechanism to capture relevant contextual information within prompts and follows a generation-validation-repair mechanism to refine and correct errors in generated unit tests.

Another area of focus is to use LLMs for property-based or specialized testing. Vikram et al. \cite{vikram_2023_can} used LLMs to generate property-based tests with API documentation. Koziolek et al. \cite{koziolek_2024_automated} studied the use of LLMs to automate test case generation for Programmable Logic Controller (PLC) and Distributed Control System (DCS) control logic. Plein et al. \cite{plein_2024_automatic}  investigated the use of LLMs, including ChatGPT and CodeGPT, for creating executable test cases from natural language bug reports. Their approach showed potential for enhancing tasks like fault localization and patch validation, with test cases successfully generated for up to 50\% of examined bugs. Wang et al. \cite{wang_2019_automatic} proposed UMTG, an NLP-based method to automate the generation of acceptance test cases from natural language use case specifications.

\subsection{Benchmark for Evaluating LLMs}

Research on creating benchmark datasets for LLMs to generate test cases is limited. One recent proposal, TESTEVAL \cite{wang_2024_testeval}, collected 210 Python programs from the online coding site LeetCodes. However, the benchmark is limited to the collected Python programs which restrict its generalization or extension to other programming languages or use patterns. To better understand how to create benchmarks for LLM’s capabilities in generating code-related artifacts, we explored two close fields: text-to-SQL translation and code generation. 

Li et al. \cite{li_2023_can} in 2023 introduced the BIRD dataset, a large-scale benchmark designed to assess LLMs' ability to parse text-to-SQL queries across diverse domains. Similarly, Wang et al. (2022) \cite{lan_2023_unite} developed the UNITE benchmark, consolidates 18 publicly available text-to-SQL datasets, creating a unified resource for evaluation. UNITE comprises over 120,000 examples from more than 12 domains, 29,000 databases, and 3,900 SQL query patterns. 

For benchmarking code generation and evaluation, HumanEval, proposed by Chen et al. in 2021 \cite{markichengchen_2021_evaluating}, was a carefully crafted benchmark consisting of 164 programming challenges. Each challenge includes a function signature, docstring, body, and multiple unit tests. Liu et al. \cite{liu_2023_is} addressed the limitations of existing benchmarks in evaluating the functional correctness of code generated by LLMs, such as ChatGPT. They identified that many benchmarks rely on insufficient and low-quality test cases, leading to an overestimation of LLM performance. To tackle this, they introduce EvalPlus, a framework that augments a significant number of additional test cases using both LLM-based and mutation-based strategies. Yan et al. \cite{yan_2023_codescope} presented CodeScope, a comprehensive benchmark for evaluating LLMs in code understanding and generation across multilingual, multi-task, and multidimensional scenarios. CodeScope covered 43 programming languages and eight tasks, including code summarization, repair, and optimization.

\subsection{Summary}

Past research highlights the importance of benchmark datasets for training and evaluating LLM’s abilities to generate computing artifacts. However, existing benchmark dataset for test case generation is highly limited, lacking flexibility for extension to other programming languages or use cases. Most existing LLM-based test generation techniques focus on prompt crafting and additional documents or data for generating tests. Additionally, they were also domain-specific and lack generalizable evaluation methods.

This raise to three key questions: \\
1. Can we develop a flexible benchmark dataset? \\
2. Can users only use a single prompt with merely source codes to LLM to generate test cases? \\
3. How well can LLMs generate test cases while inputting solely source codes?

To address these gaps, there is a need for a new, more adaptable dataset. To simulate real-world scenarios where users enter codes directly, without supplementary data like documentation, and request LLM’s test case generation though a single prompt. To achieve this, we propose GBCV, an approach designed to create is extendable benchmark dataset and facilitate studies on LLM’s test case generation ability with minimal input.

\section{Method}
Our GBCV approach begins by using fundamental Control Flow Graph (CFG) structures, integrated with data-flow analysis techniques, specifically focusing on predicate uses (p-uses) and computation uses (c-uses) for variable usage. By combining CFG with data-flow analysis, we generate diverse program variants that contain different structures and variable integrations. This systematic process allows us to create a comprehensive dataset of generated programs designed to cover multiple execution paths, value conditions, and comparisons, which serves as a robust benchmark for evaluating LLM-generated test cases.

\subsection{Overall Process}
\begin{figure}
  \centering
  \includegraphics[width=1.0\columnwidth]{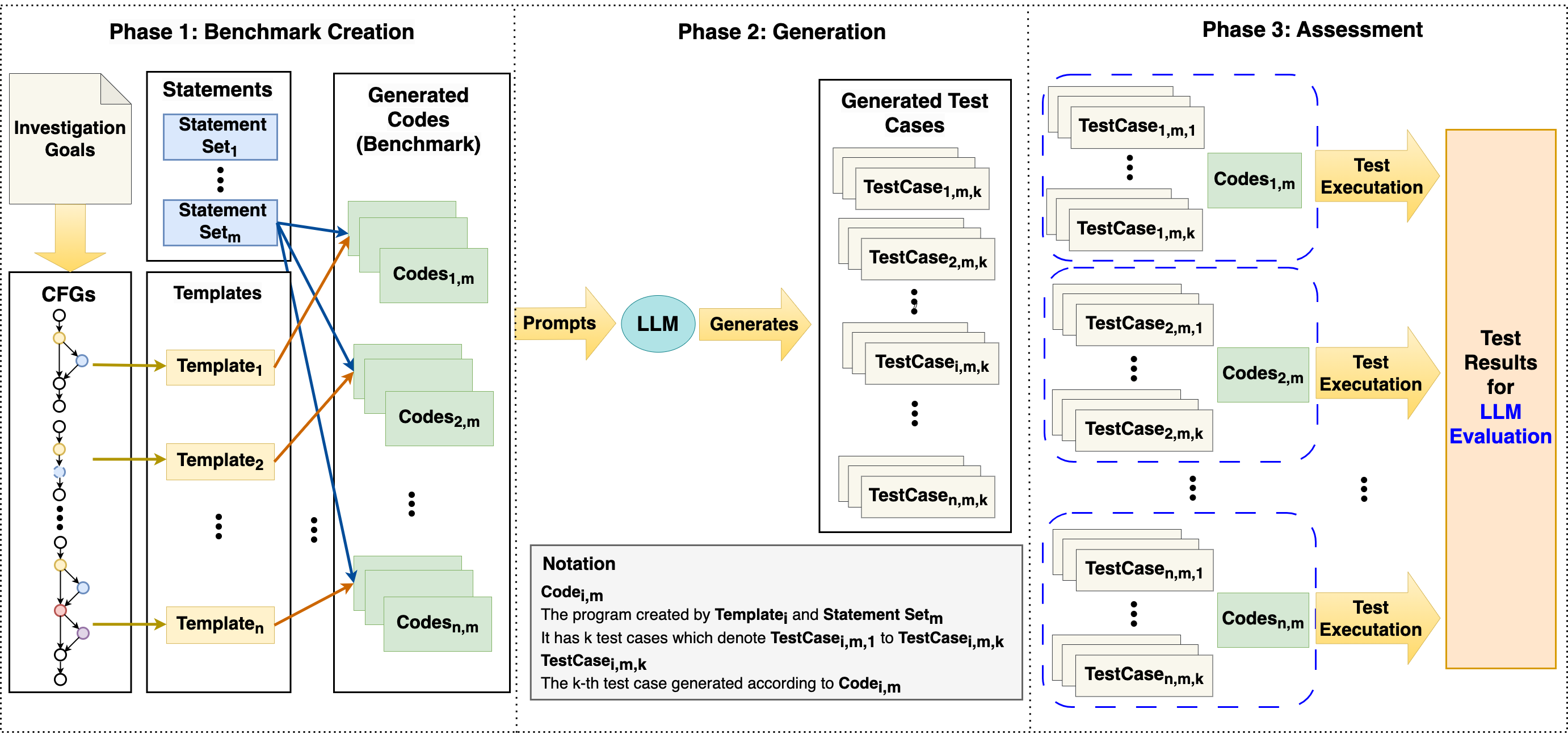}
  \caption{The process of the GBCV approach}
  \label{fig:overall_process}
\end{figure}

The GBCV approach follows a systematic procedure to automatically generate a set of programs, served as a benchmark used to evaluate the test case generation capabilities of LLMs. The process consists of three main phases (see \autoref{fig:overall_process}). The first phase focuses on benchmark creation, where programs (code) are generated to form a benchmark dataset used to prompt an LLM to generate test cases. The second phase is for test case generation. Each prompt is used to request the LLM to generate test cases. In last phase, the generated test cases (e.g., $TestCase_{i,m,k}$ in \autoref{fig:overall_process}) are used to test the corresponding programs (e.g., $Codes_{i,m}$ in \autoref{fig:overall_process}) with the expectation that the program will pass all the test cases. In this way, we can investigate whether the LLM can produce valid and correct test cases.  

The GBCV approach begins by specifying the basic program structures, such as branches and sequences, which serve as a building block to form a program. The CFG specifies the template which will be filled by partial or complete statements for the code generation. The basic structures can be incrementally added to build up into more complex combinations based on the desired level of investigation, which is a stopping point of structure creation. 

\begin{figure}
  \centering
  \includegraphics[width=1.0\columnwidth]{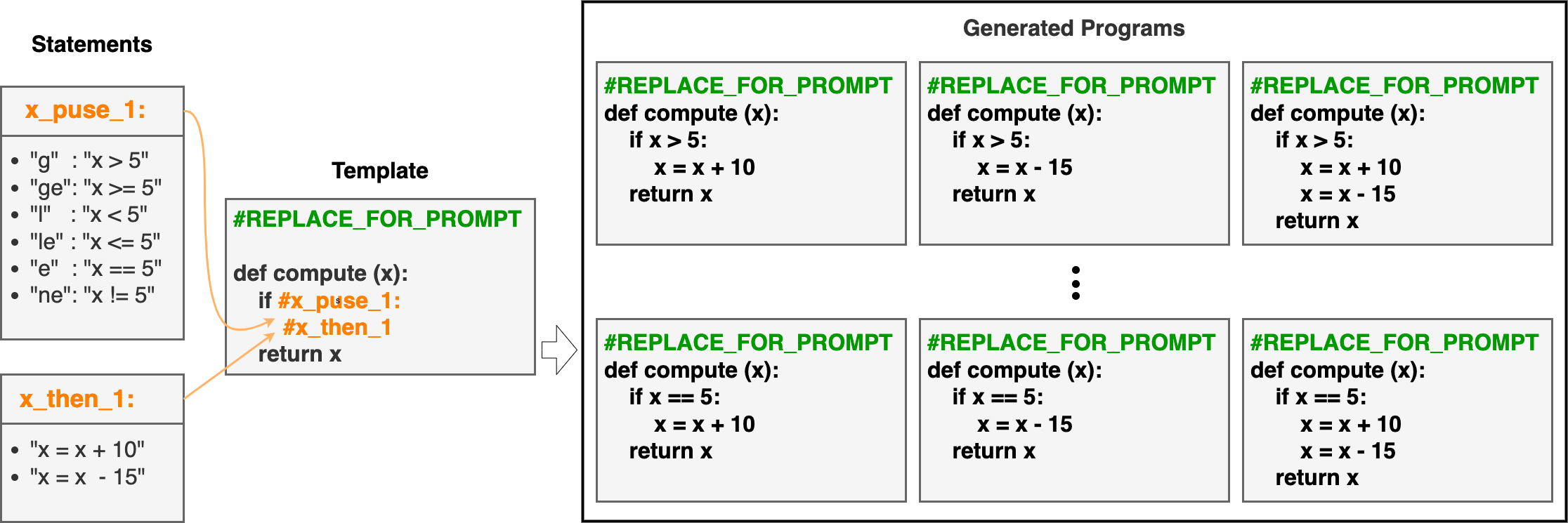}
  \caption{An example to describe the steps of the Python program generation}
  \label{fig:steps_of_py}
\end{figure}

Within the templates, placeholders are replaced with actual statements (e.g., c-use) or partial statements (e.g., p-use) to generate a complete program (see \autoref{fig:steps_of_py}). The p-use and c-use statements are designed based on several factors: the number of computations, the types of computations, and the specific decisions in a branch the evaluators plan to assess. They help in understanding how variables are utilized within the program's logic. One can use a complicated combination of conditions to test LLM’s logical reasoning capabilities. Hence, adding statements also increase the overall complexity of the programs.

When specifying templates and statements, language-specific details must be considered. For example, ava and Python use different syntax for combining predicates. Java employs x>4 || x <10, while Python uses x>4 or x<10. Therefore, to apply GBCV across different languages, the same abstractions - CFG and placeholder locations can be reused, but language-specific details are put while formatting the template and statements. 

Finally, to assess the LLM’s test case generation, the metrics are calculated against the test outputs across different types of structure and computations. The users can ensure a comprehensive understanding of the model's capabilities and limitations.

\subsection{Guiding Principles of Benchmark Creation}

The GBCV is a bottom-up approach to generate programs for assessing the LLM. Several factors decide the final generated programs. We describe them in the following \autoref{tab:Table_guiding_principles}.

\begin{table}[]
\caption{Percentage of untestable program in every category.}
\centering
\begin{tabular}{llll}
\toprule
  Principles     & Element Description & \begin{tabular}{@{}c@{}}Complexity \\ Level*\end{tabular}   & Examples \\
\midrule
\multirow{7}{*}{\parbox{4cm}{Types of predicates with number of predicates}} & Single comparison: equality & L & x == 18 \\
     & \parbox[t]{4.5cm}{Single comparison: relational} & L & x > 18 \\
     & \parbox[t]{4.5cm}{Single predicate with computation} & M & \begin{tabular}{@{}l@{}}is\_postive(x) \\ x\%2 == 0 \end{tabular}  \\
     & \parbox[t]{4.5cm}{Compound predicates (two predicates)}  & M  & \begin{tabular}{@{}c@{}}x >= 18 and y < 5 \\ x >= 18 \&\& y < 5\end{tabular}   \\
     & \parbox[t]{4.5cm}{Compound predicates (multiple predicates)} & M-H   &  \begin{tabular}{@{}c@{}} x >= 18 and y < 5 and z \\ x >= 18 \&\& y < 5 \&\& z\end{tabular}    \\     
     & \parbox[t]{4.5cm}{Compound predicates (two predicates) with computations)}  & M-H   & \begin{tabular}{@{}c@{}} x+ y > 5 and x\%2 == 0 \\ x+ y > 5 \&\& x\%2 == 0\end{tabular}    \\  
     & \parbox[t]{4.5cm}{Compound predicates (multiple predicates) with computations}  &  H  & \begin{tabular}{@{}c@{}} x+ y > 5 and x\%2 == 0 and z \\ x+ y > 5 \&\& x\%2 == 0 \&\& z\end{tabular}    \\  
\midrule
\multirow{2}{*}{\parbox{4cm}{Number of computations}} & 1-3 & L &  \\
     & \parbox[t]{4.5cm}{>3} & M-H &  \\
\midrule
\multirow{2}{*}{\parbox{4cm}{Type of computations}} & \parbox[t]{4.5cm}{Only four simple four simple arithmetic operations} & L & x = x + 5 \\
     & \parbox[T]{4.5cm}{Beyond four simple four simple arithmetic operations} & M-H & \begin{tabular}{@{}l@{}} math.sqrt(5) \\ 5**2  \end{tabular}  \\

\midrule
\multirow{5}{*}{\parbox{4cm}{Variable datatypes or constant values or used values}} & \parbox[t]{4.5cm}{Integer} & L & 25 \\
     & \parbox[T]{4.5cm}{Floating point} & L-M & 1.22  \\
     & \parbox[T]{4.5cm}{String} & L-M & “abc”  \\
     & \parbox[T]{4.5cm}{Boolean} & L & True/False  \\
     & \parbox[T]{4.5cm}{Complex} & M-H & \begin{tabular}{@{}l@{}} Object \\ Nested list  \end{tabular}  \\

\midrule
\multirow{3}{*}{\parbox{4cm}{Number of basic structures}} & \parbox[t]{4.5cm}{1} & L &  \\
     & \parbox[T]{4.5cm}{2} & L-M &  \\
     & \parbox[T]{4.5cm}{>2} & M-H &   \\

\midrule
\multirow{4}{*}{\parbox{4cm}{Formation of multiple basic structures}} & \parbox[t]{4.5cm}{Sequential} & L-M &  \\
     & \parbox[T]{4.5cm}{Nested} & M-H &  \\
     & \parbox[T]{4.5cm}{Recursive} & H &   \\
     & \parbox[T]{4.5cm}{Mixed} & H &   \\
     
\midrule
\multicolumn{4}{l}{*L: Low, M: Middle, H: High} \\

\bottomrule
\end{tabular}
\label{tab:Table_guiding_principles}
\end{table}

The complexity levels leverage a combination of cyclomatic complexity \cite{watson_1996_structured}, cognitive complexity \cite{jingqiushao_2003_a, misra_2006_a}, and the reasoning behaviors of the LLM. Both cyclomatic and cognitive complexity account for the code structure (e.g., CFG) and decision points. Specifically, cyclomatic complexity is categorized into four levels: low (1–4), moderate (5–7), high (8–10), and very high (above 10, typically recommended for refactoring). Unlike cyclomatic complexity, cognitive complexity differentiates predicates and program structures and considers logical operators in evaluation. When assessing an LLM’s reasoning behaviors, we consider factors such as the number of statements, predicates, and datatypes. The considerations about the number of statements and predicates are drawn from prior research on chain-of-thought reasoning \cite{jasonzhanshunwei_2022_chainofthought}. The datatype selection leverages the insights from previous studies on type-aware reasoning \cite{liu_2023_is, winterer_2020_on}.  

Once the complexity level and the specific LLM behaviors of interest are identified, the appropriate CFG and statements can be selected accordingly. The selection also depends on the intended coverage range. For instance, if the objective is to assess whether an LLM can recognize a sequence of computations, incorporating multiple c-use nodes is necessary. Likewise, if datatype awareness is important, incorporating diverse or complex datatypes should be considered.

\subsection{Dataset Creation}
Since the programs in the benchmark dataset for evaluating LLMs' test case generation ability are created based on the guiding principles of the GBCV approach, the number and types of generated programs dynamically adjust according to the user's specified assessment goals. Given that the primary objective of this study is to introduce the GBCV approach and gain an initial understanding of LLMs' test case generation capabilities as a foundation for future research on more complex or real-world programs, our assessments focus on relatively simple structures, data types, and arithmetic computations. Therefore, we intentionally select structures, data types, and computation levels between low to middle complexity.

\begin{figure}[hbt!]
  \centering
  \includegraphics[width=1.0\columnwidth]{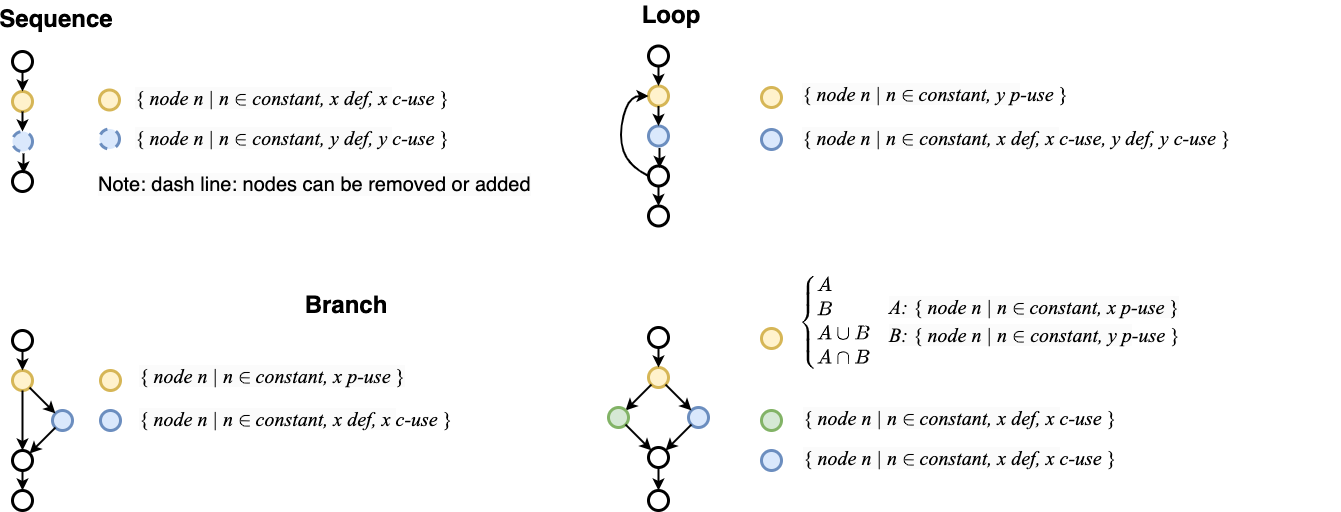}
  \caption{Simple program structure for code generation. The colors point to placeholder nodes that are replaced with the statements from corresponding sets of statement candidates. Examples include x def (x=15), x def with x c-use (x=x+10), y def (y = 7), and y-def with y c-use (y = y + 7) in Sequence. These example statements are used to replace the respective nodes.}
  \label{fig:simple_prog_struct}
\end{figure}

\begin{figure}[hbt!]
  \centering
  \includegraphics[width=1.0\columnwidth]{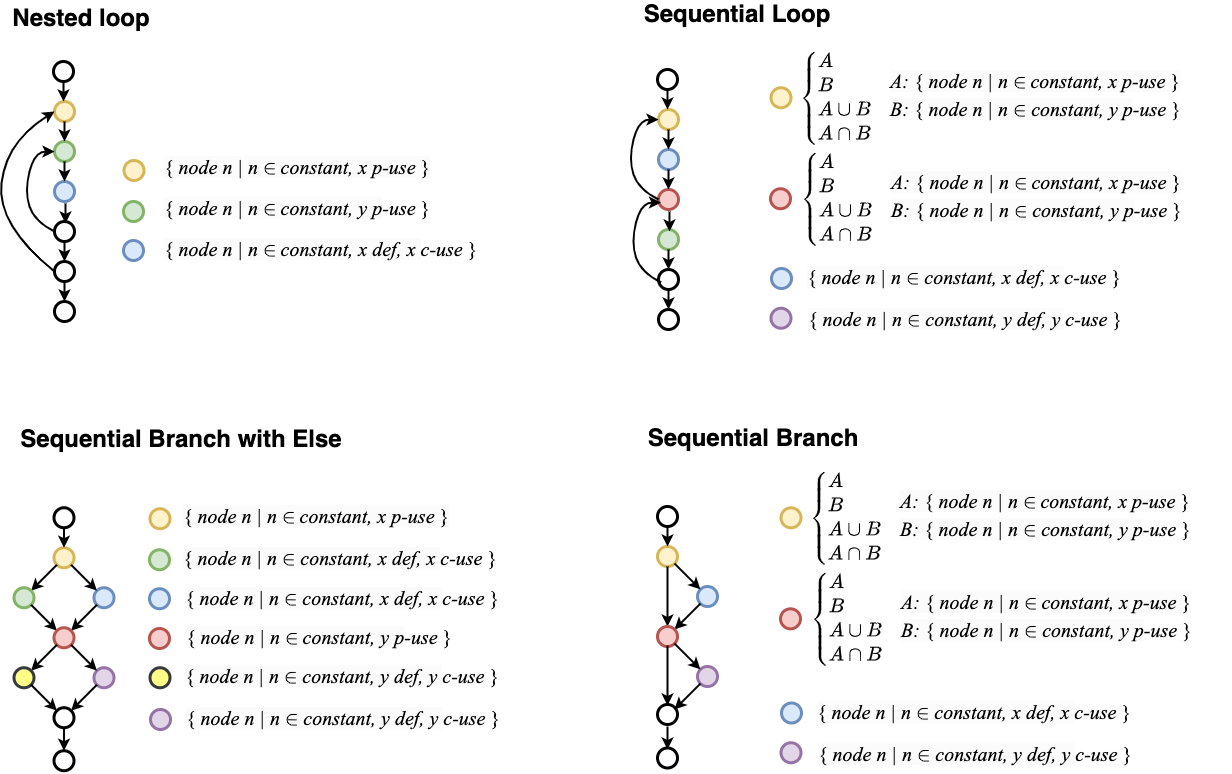}
  \caption{Composite program structure for code generation. The color nodes indicate placeholders that are replaced with statements selected from the corresponding candidate sets. In addition to x def, y def, x c-use, and y c-use described earlier, examples of x p-use and y p-use include x < 5 and y > 10, respectively. Compound predicates consider union (or) and intersection (and) in the logical operations.}
  \label{fig:composite_prog_struct}
\end{figure}

The following discusses the investigation goals in this study. \autoref{fig:simple_prog_struct} and \autoref{fig:composite_prog_struct} illustrate the control flow structures and set of nodes defined we used based on our investigation goals.

\begin{enumerate}
\item Boundary and comparison \\
This evaluation assesses the LLM's ability to perform logical operations and comparisons. The p-use of a variable in a conditional statement, such as 'x > 10', checks whether the LLM can correctly identify the boundary value (i.e., 10), resulting in true/false branches when the comparison clause contains constant values. Successful boundary value detection has two key characteristics: First, the input value chosen in a test case matches the boundary value (e.g., an input value of 10 for 'x > 10'). Second, the chosen input values can lead to both true and false execution paths, depending on the comparison.

Therefore, according to the low to middle complexity levels of the type of predicates and numbers of predicates, and low complexity for datatype, we use two variables to form compound predicates and the datatypes and values for comparison are integers.

In the branch structure (see \autoref{fig:simple_prog_struct}), the value we chose for the variable x in the comparison predicate is 5, to maintain the consistency for investigation, the integer value 5 is also used in the compound predicate in other branch-like structure (see \autoref{fig:composite_prog_struct}). We want to examine if integer value (boundary) can be successfully detected given various structures, comparisons and logical operations. Therefore, the following list shows some examples of the x p-use and y p-use nodes we used.

        \begin{itemize}
          \item Branch, Sequential Branch, Sequential Branch with Else:
          \begin{itemize}
            \item x p-use set 1: x > 5, x < 5 …, x == 5
            \item x p-use set 2: x > 15, x < 15 …, x == 15
            \item y p-use: y > 10, y < 10 …, y == 10
            \item Compound predicate for x and y: x > 5 and y > 10, … x == 5 or y == 10
          \end{itemize}
        \end{itemize}
  
\item Computation \\
This evaluation examines how well the LLM can perform computations involving variables. The c-use nodes in a sequence test the LLM's ability to compute results, particularly when operations involve multiple variables, such as 'x' and 'y'. This helps assess the LLM's effectiveness in handling simple arithmetic operations. 

Therefore, those def and c-use in all the structures present the computation. To note that, not all the programs contain computation. For example, a program in the Branch can just change the definition of the variable. The following list some examples of computation used in program generation.

        \begin{itemize}
          \item Sequence, Loop, Branch, Sequential Branch, Sequential Branch with Else:
          \begin{itemize}
            \item x def with x c-use: x=x+10, …x=x*7, x=x-7
            \item y def with y c-use: y=y+7, …, y=y-7
            \item x def with x c-use and y c-use: x=x+y+10, …, x=x+y-7
          \end{itemize}
        \end{itemize}

When examining the correct test case generation in the case of computations joining with branch in Branch-like structure, we further want to investigate if the LLM can reason both computation and boundary correctly without interfering each other. Similarly, the computations with loops test whether can remember the temporary results during the iterations. 

\item Iteration \\
Simple loops are used to evaluate the LLM's ability to determine the correct number of iterations and accurately iterate through a portion of the code. This scenario is more complex than branches or sequences, as it may require han-dling a combination of computations and predicate uses. The following list the iteration conditions we used in the Loop-like structures. By comparing with constant number of loops and dynamic number of loops (i.e., number of iteration depending on the input, whether the LLM detect the differences can be investigated.

        \begin{itemize}
          \item Loop, Sequential Loop, Nested Loop:
          \begin{itemize}
            \item Constant value: 3
            \item Dynamic value: y p-use
          \end{itemize}
        \end{itemize}
        
\end{enumerate}

Complex control flow structures are composed of multiple basic structures. They are used to evaluate how well the LLM can manage combined operations during test case generation. For example, evaluating two branches can reveal whether the LLM can correctly handle multiple branching conditions. By assessing these combinations, we can also gain a deeper understanding of the LLM's ability to manage boundary detection, computations, and iterations effectively, as well as identify the impact of new combinational effects within complex structures.

Two variables are used for examining test case generation under multi-variable conditions in our study. However, if further exploration is needed on how LLMs handle additional variables, our approach allows the definition of more variables in the control flow structure to generate additional programs for testing. The number of created programs depends on the combination of candidates used to replace the placeholders. Programs with the same control flow structure but different predicates and computations can let us understand which types of usage or structures can be the potential weakness in LLM’s test case generation abilities. 

To facilitate prompting, we utilize a placeholder, \#REPLACE\_FOR\_PROMPT, in the template. In our study, the prompting placeholder is replaced with the instruction: "Given the program, please create test cases that can pass 100\% statement coverage". While achieving 100\% statement coverage is not the primary objective of this study, this prompt encourages the LLM to generate a diverse set of test cases rather than a single one. Moreover, our aim is to evaluate the LLM’s ability to generate correct test cases on its first attempt without iterative refinements on prompts or follow-up questions. Thus, the request for full statement coverage serves merely as a mechanism to elicit comprehensive output and does not impact the LLM's inherent capability of generating accurate test cases. 

\subsection{Metrics}
To define our metrics, we first need to introduce the following terminologies, which describe several characteristics of the test cases generated by LLMs generated.

\begin{enumerate}
    \item Complete (valid) and incomplete (invalid) test case \\
    For a test case in the LLM's response to be valid, it must be complete, meaning it includes both inputs and expected results. If a generated test case does not contain expected results, it is considered incomplete, therefore, invalid.

    \item Correct test case \\
    A correct test case must meet two conditions. First, it must be a complete test case, meaning it includes inputs and expected outputs. Second, when applied to test the generated program, the actual output must match the expected output. If an LLM cannot produce correct test cases, it indicates poor test case generation capability.
    
    \item Untestable program (codes) \\
    The untestable program in this paper does not mean that the program itself cannot be tested. Rather, it refers to a generated program that lacks valid generated test cases for its testing. Therefore, let $T$ be the set of all LLM generated test cases for a generated program P after one prompt request and $T = \{T_{1}, T_{2}, … ,T_{n}\}$. If all elements in $T$ are invalid test cases, the program $P$ cannot further be tested. However, if at least one valid test case $T_{i}$ exists, the program $P$ remains testable, as it can still be tested by $T_{i}$. In this case, we check if the valid test cases are correct and classify invalid test cases as incorrect test cases. The untestable program is special when we investigate the LLM’s ability of test case generation. It indicates that the LLM cannot even produce a valid test case that can be used to test the program.

\end{enumerate}

There are multiple test cases generated by the LLM after a prompt request is associated with the program $P$. Hence, the error rate $E_{p}$ of the LLM test case generation associated with program $P$ is calculated as Equation (1), where $N_{t,p}$ represents the total number of generated test cases according to a program and $N_{s,p}$ is the number of correct test cases. The error rate helps identify weaknesses in the LLM's ability to handle specific programs related to its structure, datatype, predicate, or computations.

\begin{equation}
E_{p}=(N_{t,p}-N_{s,p})/N_{t,p}
\end{equation}

Therefore, the average error rate $AvgE_{i}$ is calculated by dividing the sum of the error rate per program by the total number of program $TP_{i}$ in the category $i$, as shown in the Equation (2).   

\begin{equation}
AvgE_{i}= \sum E_{p}/TP_{i}
\end{equation}

The untestable program rate can be determined within a category. Equation (3) calculates $R_{i}$, the percentage of untestable programs within each program structure category. In this equation, $W_{i}$ represents the total number of untestable programs in code structure category $i$ and $TP_{i}$ is the total number of programs in that category $i$. 

\begin{equation}
R_{i}= W_{i}/TP_{i}
\end{equation}

The incomplete test case rate based on an LLM model also implies level of understanding of the LLM on the definition of a complete (valid) test case. Equation (4) shows how to calculate this rate, where $TR_{i}$ is the incomplete test case rate, $IT_{i}$ represents the number of all the incomplete test cases generated by the LLM model $i$, $CT_{i}$ is the number of all the complete test cases generated by the LLM model $i$, and $TT_{i}$ is the number of all the test cases generated by the model $i$ (i.e., the sum of the total test cases generated per category by the model $i$)

\begin{equation}
TR_{i}= IT_{i}/TT_{i} = (TT_{i} - CT_{i})/TT_{i}  
\end{equation}

Each program does not have equal number of test cases; therefore, the average test case number $AvgTN_{i}$ within the category $i$ is calculated by dividing the total generated test cases $TN_{i}$ in the category $i$ by the total number programs $TP_{i}$ in the category $i$, as shown in Equation (5).

\begin{equation}
AvgTN_{i}= TN_{i}/TP_{i}
\end{equation}

\section{Result and Discussion}
\subsection{Result}
The dataset we used is not extracted from any existing repository. A total of 786 Python programs were all generated using seven specified types of control flow structures: Branch, Loop, Nested Loop, Sequence, Sequential Branch, Sequential Branch with Else, and Sequential Loop. \autoref{tab:avg_err_rate_across} presents the average complexity of the generated programs within a category by using the average source lines of code (SLOC) and illustrates the coverage of each program type as a percentage of the total generated programs. The program that includes at most two variables can be used to assess the models' performance in handling multiple predicates or computation. Sequential Branch and Sequential Branch with Else contains various combinations of c-use nodes and compound predicates, their coverages are relative higher than other categories. To simulate how regular users use prompts to ask for test cases for a program from an LLM as the initial investigation done by our generated benchmark, we used the generated programs to evaluate three language models: GPT-3-Turbo, GPT-4o-mini, and GPT-4o, rather than LLMs specifically trained for coding.

\begin{table}
 \caption{Average complexity and coverage across categories.}
  \centering
  \begin{tabular}{lll}
    \toprule
    Category     & Average Complexity (Average SLOC) & Coverage (\% of Total Programs) \\
    \midrule
    Branch       & 6 & 9.16\%   \\
    Loop         & 5 & 3.94\%   \\
    Nested Loop  & 6 & 3.81\%   \\
    Sequence                     & 4.4 & 2.22\%   \\
    Sequential Branch            & 8   & 37.1\%   \\
    Sequential Branch with Else  & 12  & 37.15\%   \\
    Sequential Loop              & 7   & 6.62\%   \\
    \bottomrule
  \end{tabular}
  \label{tab:avg_err_rate_across}
\end{table}

\autoref{tab:inc_tc_rate} presents the evaluation results, including the total number of test cases generated by each model, the number of complete test cases, and the percentage of incomplete test cases calculated by Equation (4). Among the models, GPT-3-Turbo had the highest rate of incomplete test cases at 32.74\%, making it the least effective. GPT-4o-mini performed the best, with an incomplete test case rate of 6.1\%, while GPT-4o had a similar performance, with an incomplete rate of 7.5\%. 

\begin{table}
 \caption{Incomplete test case rates across three models.}
  \centering
  \begin{tabular}{llll}
    \toprule
    Model i     & Total Test Cases $TT_{i}$  & Complete Test Cases $CT_{i}$  & Incomplete Test Case Rate $TR_{i}$ \\
    \midrule
    GPT-3-Turbo   & 2941 &  1978  & 32.74\%   \\
    GPT-4o-mini   & 3030 &  2845  & 6.1\%   \\
    GPT-4o        & 3032 &  2803  & 7.5\%   \\
    \bottomrule
  \end{tabular}
  \label{tab:inc_tc_rate}
\end{table}

\begin{figure}
  \centering
  \includegraphics[width=1.0\columnwidth]{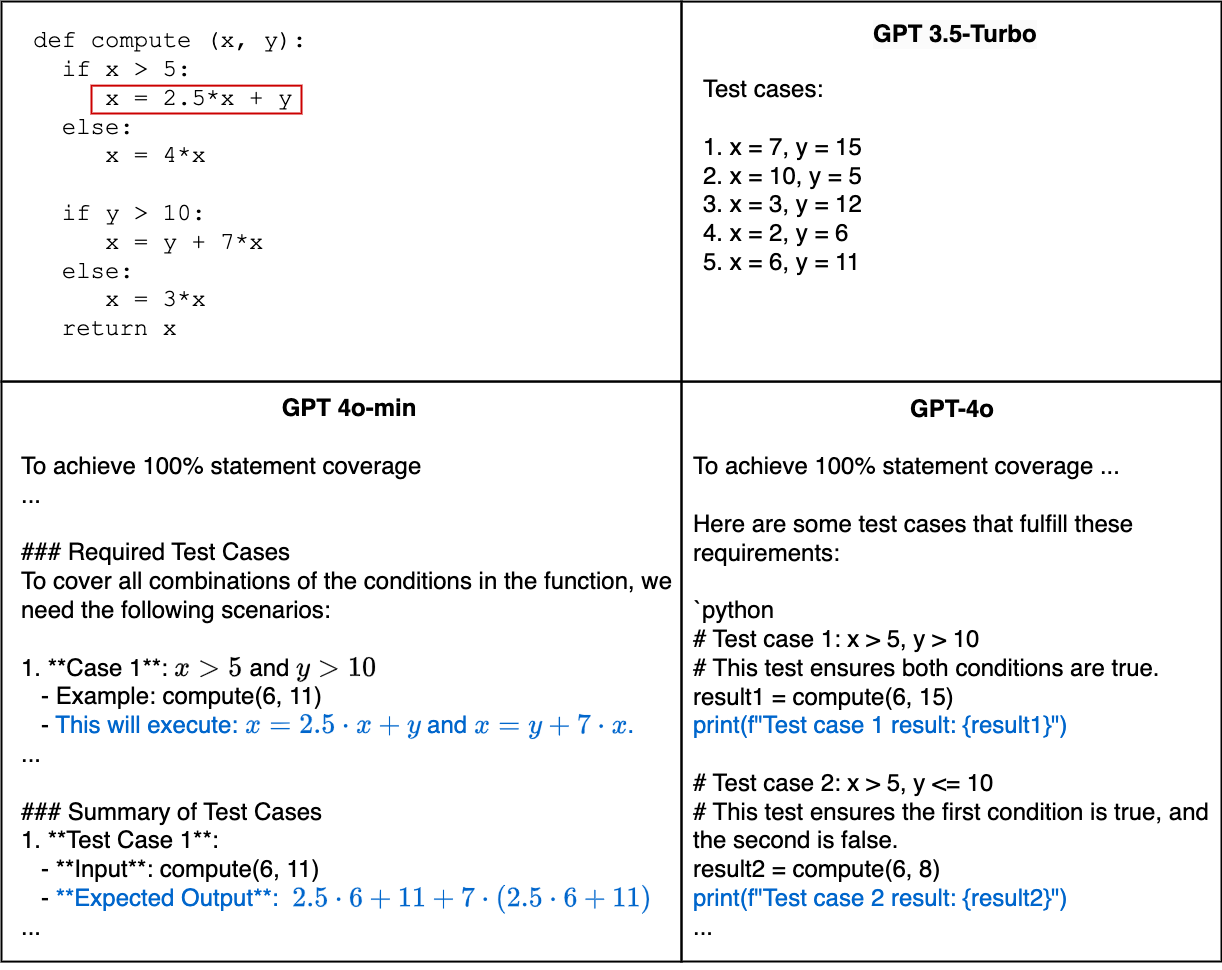}
  \caption{Examples of incomplete test cases - program that do not provide expected output values in the response for all three LLMs }
  \label{fig:incomplete_tcs}
\end{figure}

\begin{figure}
  \centering
  \includegraphics[width=0.7\columnwidth]{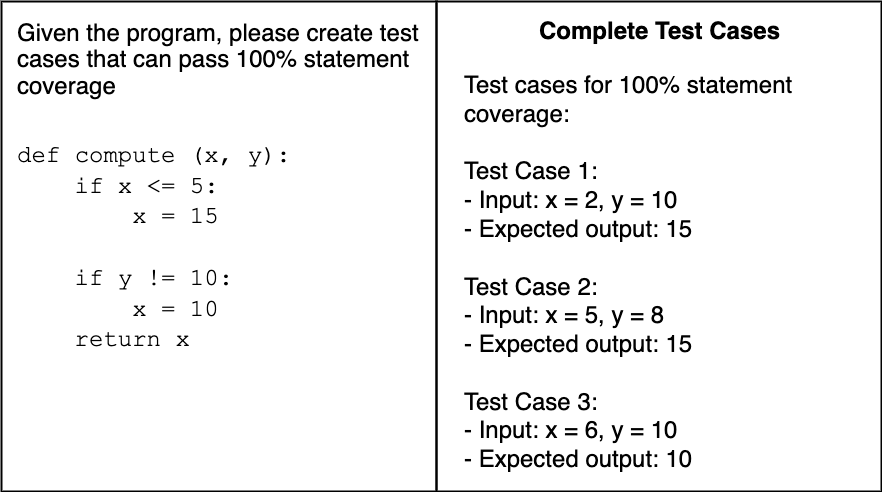}
  \caption{An example of complete test cases generated by GPT-3.5-Turbo. The response does not contain any explanations besides inputs and expected outputs.}
  \label{fig:complete_tcs}
\end{figure}

\begin{figure}
  \centering
  \includegraphics[width=1.0\columnwidth]{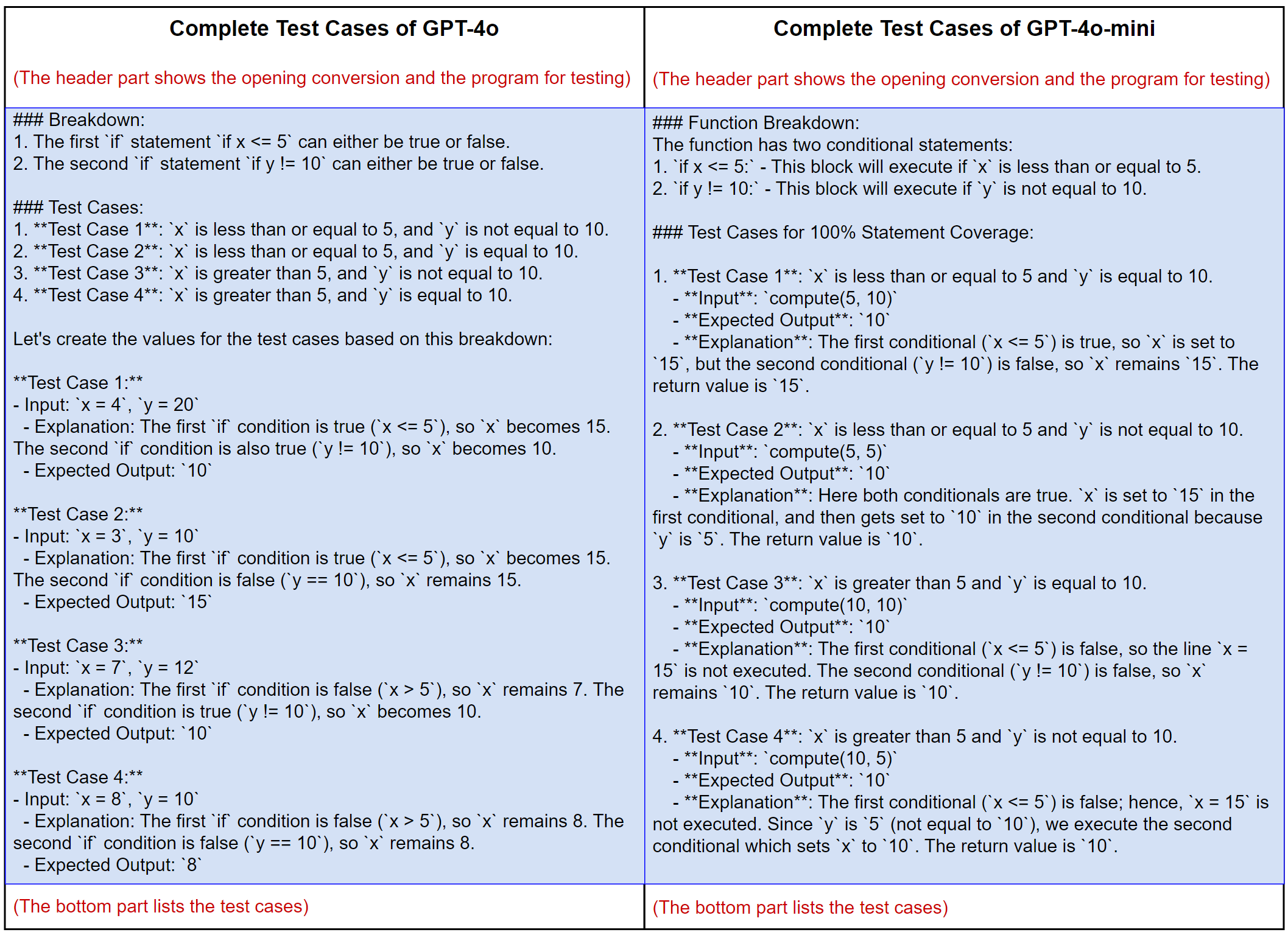}
  \caption{An example response of complete test cases with detailed explanations generated by GPT-4o fam-ily for the same program as in Figure 6.}
  \label{fig:correct_incorrect}
\end{figure}

\autoref{fig:incomplete_tcs} illustrates examples of incomplete test cases, while \autoref{fig:complete_tcs} and \autoref{fig:correct_incorrect} present examples of a complete test case. In the case of GPT-3-Turbo, the generated responses often lacked the detailed explanations that were included in the outputs of GPT-4o and GPT-4o-mini. These detailed explanations are step-by-step descriptions on how expected outputs are derived in GPT-4o and GPT-4o-mini (see Figure 7). Specifically, GPT-4o and GPT-4o-mini provided expected test results in the form of either repeating the function inputs or elaborating on the computations involving those inputs (see the sentence in blue in Figure 5). However, these elaborations were often inaccurate, highlighting the need for further improvement in the models' ability to provide reliable and precise explanations alongside the generated test cases.

\autoref{tab:three_performance} summarizes the performance of three LLMs, GPT-3-Turbo, GPT-4o-mini, and GPT-4o, across various categories of control flow structures, showing their average error rates and number of test cases. They are calculated by Equation (2) and Equation (5), respectively. It is important to note that the evaluation only considered complete test cases generated by each model. As expected, GPT-4o and GPT-4o-mini outperformed GPT-3-Turbo in most categories, except for the “Loop” category. GPT-3-Turbo struggled significantly, with error rates exceeding 80\%, in the "Nested Loop," "Sequential Branch with Else," and "Sequential Loop" categories. GPT-4o-mini had the lowest performance in the "Nested Loop" category but achieved its best results in the "Sequential" category. Interestingly, GPT-4o face challenges in the "Loop" category, while excelling in the "Branch" category, matching the top performance of GPT-4o-mini in this area.

\begin{table}[]
\caption{Average test cases and error rate across three models.}
\centering
\begin{tabular}{llll}
\toprule
  Category     & Models & Average Test Cases ($AveTN_{i}$)  & Average Error Rate ($AvgE_{i}$) \\
\midrule
\multirow{3}{*}{Branch} & GPT-3-Turbo & 3.62 & 0.196  \\
                      & GPT-4o-mini & 3.5  & 0.02   \\
                      & GPT-4o      & 2.92 & 0.058  \\
\multirow{3}{*}{Loop}  & GPT-3-Turbo & 4    & 0.5  \\
                        & GPT-4o-mini & 4.67 & 0.267   \\
                        & GPT-4o      & 3.33 & 0.6  \\  
\multirow{3}{*}{Nested Loop}    & GPT-3-Turbo & 3    & 0.826  \\
                                & GPT-4o-mini & 3.64 & 0.511   \\
                                & GPT-4o      & 3.5  & 0.363  \\ 
\multirow{3}{*}{Sequence}   & GPT-3-Turbo & 3    & 0.2  \\
                            & GPT-4o-mini & 3.75 & 0.1   \\
                            & GPT-4o      & 4.67 & 0.167  \\ 
\multirow{3}{*}{Sequential Branch} & GPT-3-Turbo & 3.69 & 0.587  \\
                                    & GPT-4o-mini & 3.79 & 0.154   \\
                                    & GPT-4o      & 4.01 & 0.084  \\ 
\multirow{3}{*}{Sequential Branch with Else}   & GPT-3-Turbo & 3.89 & 0.882  \\
                                                & GPT-4o-mini & 4.08 & 0.224  \\
                                                & GPT-4o      & 3.98 & 0.196  \\ 
\multirow{3}{*}{Sequential Loop}  & GPT-3-Turbo & 3.21 & 0.883  \\
                                    & GPT-4o-mini & 3.91 & 0.506   \\
                                    & GPT-4o      & 3.13 & 0.290  \\ 
\bottomrule
\end{tabular}
\label{tab:three_performance}
\end{table}

\autoref{tab:untestable_programs} shows the percentage of untestable programs in each category. GPT-3-Turbo had a high percentage of untestable programs in every category (all above 14.29\%). GPT-4o-mini performed comparably the best, though GPT-4o had similar performance as GPT-4o-mini. A lower percentage of untestable programs indicates that the LLMs were able to provide complete test cases for those programs; however, this does not guarantee that the generated test cases are all correct. This suggests that some complete test cases may have been generated inaccurately through additional improper reasoning.

\begin{table}[]
\caption{Percentage of untestable programs in every category.}
\centering
\begin{tabular}{llll}
\toprule
  Category     & Models & \% of Untestable Programs* \\
\midrule
\multirow{3}{*}{Branch} & GPT-3-Turbo & 19.4\%   \\
                        & GPT-4o-mini & 1.38\%   \\
                        & GPT-4o      & 6.94\%   \\
\multirow{3}{*}{Loop}   & GPT-3-Turbo & 33.3\%   \\
                        & GPT-4o-mini & 0\%      \\
                        & GPT-4o      & 0\%      \\  
\multirow{3}{*}{Nested Loop}    & GPT-3-Turbo & 14.29\%     \\
                                & GPT-4o-mini & 0\%         \\
                                & GPT-4o      & 14.29\%     \\ 
\multirow{3}{*}{Sequence}   & GPT-3-Turbo & 0\%    \\
                            & GPT-4o-mini & 0\%    \\
                            & GPT-4o      & 20\%   \\ 
\multirow{3}{*}{Sequential Branch} & GPT-3-Turbo  & 31.77\%   \\
                                    & GPT-4o-mini & 2.43\%    \\
                                    & GPT-4o      & 3.3\%     \\ 
\multirow{3}{*}{Sequential Branch with Else}   & GPT-3-Turbo & 42.19\%  \\
                                               & GPT-4o-mini & 23.44\%  \\
                                               & GPT-4o      & 17.19\%  \\ 
\multirow{3}{*}{Sequential Loop}    & GPT-3-Turbo  & 26.92\%   \\
                                    & GPT-4o-mini  & 11.54\%   \\
                                    & GPT-4o       & 23.08\%   \\ 
\bottomrule
\end{tabular}
\label{tab:untestable_programs}
\end{table}

\subsection{Discussion}
Our approach effectively and automatically evaluates the LLM's test case generation capabilities based on defined metrics, providing flexibility to extend program structures from simple to complex. This flexibility allows us to discover common behaviors, such as iteration, across different program types, providing deeper insights into how LLMs handle similar operations or usage among programs with various complexities. The automated evaluation identifies areas where the models excel and where they encounter poor performance, while a qualitative assessment, done by the manual process, provides a detailed analysis of the models' outputs. This combined method ensures that we can assess both the accuracy and the reasoning behind the generated test cases, offering a comprehensive understanding of the strengths and weaknesses of different LLMs. The following sections present our findings from analyzing the responses of different language models in detail.

\subsubsection{Simple vs Composite Structure}
Our analysis revealed that GPT-4o and GPT-4o-mini exhibit different strengths when handling program structures of varying complexity. GPT-4o performs better with composite structures, whereas GPT-4o-mini excels with simpler ones (see average error rate in \autoref{tab:three_performance}). GPT-4o appears to enhance the ability to understand composite programs but tends to generate more erroneous test cases when dealing with simple structures. This is evident from the higher rate of incorrect test cases produced by GPT-4o. Furthermore, compared to GPT-4o-mini when generating test cases, GPT-4o is more likely to provide detailed explanations in its responses, which can also introduce inaccuracies if the reasoning is flawed. This suggests that the level of detail in GPT-4o’s responses may contribute to its higher error rate in simpler programs.

The high rates of untestable programs suggest that GPT-4o has challenges interpreting certain programs, resulting in situations where expected outcomes are absent. Despite these challenges, the GPT-4o and GPT-4o-mini models outperform GPT-3-Turbo, particularly when dealing with composite program structures. In the "Sequential Branch" category (see \autoref{tab:untestable_programs}), the performance gap between the GPT-4o family models and GPT-3-Turbo is particularly evident, highlighting the enhanced ability of GPT-4o family in managing complex control flows. This indicates that while GPT-4o has a strong capacity for dealing with complexity, there is still room for improvement in handling simpler scenarios and reducing the rates of untestable cases.

\subsubsection{Computation}
Doing computational tasks or problems still remains a challenge for test case generation as previous studies have reported poor interpretation on arithmetic computation in earlier GPT models, despite improvements in later versions \cite{openai_2023_gpt4}. Our evaluation found that similar computational difficulties are present in test case results for the computation nodes (c-use) within generated programs. This indicates arithmetic-related inaccuracies are still prevalent. This suggests that further refinement is needed for improvement in generating test cases for programs involving arithmetic computations. 

Specifically, programs involving the c-use of two variables, such as 'x = x + y + 5', often lead to incorrect expected results in the generated test cases, regardless of the overall program complexity. Furthermore, scenarios that include multiple c-use nodes within composite structures, such as Nested Loops or Sequential Loops, can further amplify these computational challenges during test case generation. In summary, computational challenges are common across all program types. Addressing these issues will be crucial to enhancing the overall robustness of LLM-generated test cases, particularly for programs with heavy arithmetic operations or composite structures.

\subsubsection{Boundary and Comparison}
We observed that all language models perform relatively well in detecting the boundary values of a variable within simple conditional statements (e.g., x > 5 or y < 10). LLMs were capable of selecting the appropriate integer values to identify boundaries, even when handling more complex structures. For instance, \autoref{tab:presentage_untestable} presents an example of test cases generated by all models in the Sequential Branch category. It is worth noting that many successful boundary detections occur in conditional statements that do not involve any computation (e.g., x + y > 5).

\begin{table}[]
\caption{Percentage of untestable program in every category.}
\centering
\begin{tabular}{llll}
\toprule
  Code     & GPT-3-Turbo & GPT-4o & GPT-4o-mini \\
\midrule
\multirow{8}{*}{\begin{tabular}[c]{@{}l@{}}def compute (x, y):\\ \hspace{0.3cm}if x \textgreater 5:\\  \hspace{0.6cm}x = x + y + 10\\ \\ \hspace{0.3cm} if y == 10:\\ \hspace{0.6cm} x = x + y - 7\\  \hspace{0.3cm}return x\end{tabular}} & Input 1: (x, y) = (6, 10) & Input 1: (x, y) = (6, 10) & Input 1: (x, y) = (6, 5) \\
     & Input 2: (x, y) = (7, 10) & Input 2: (x, y) = (5, 10) & Input 2: (x, y) = (5, 10) \\
     & Input 3: (x, y) = (4, 10) & Input 3: (x, y) = (6, 5) & Input 3: (x, y)  = (7, 10)   \\
     & Input 4: (x, y) = (6, 8)  & Input 4: (x, y) = (5, 5)  & Input 4: (x, y) = (4, 5)  \\
     & Input 5: (x, y) = (3, 11) &    &   \\     
     &   &    &   \\  
     &   &    &   \\  
     &   &    &   \\  
\bottomrule
\end{tabular}
\label{tab:presentage_untestable}
\end{table}

\subsubsection{Iteration}
All the language models demonstrate relatively high error rates and high untestable program rates in all the loop-like structures. These outcomes indicate that LLMs face challenges in handling iterations, particularly in programs that require continuous computation. Given that arithmetic and computation are already challenging for these models, it is not surprising to see poor performance when loops are introduced. Iterations inherently require repeated computations and maintaining state over multiple cycles, which complicates the task for language models.

\begin{figure}
  \centering
  \includegraphics[width=0.75\columnwidth]{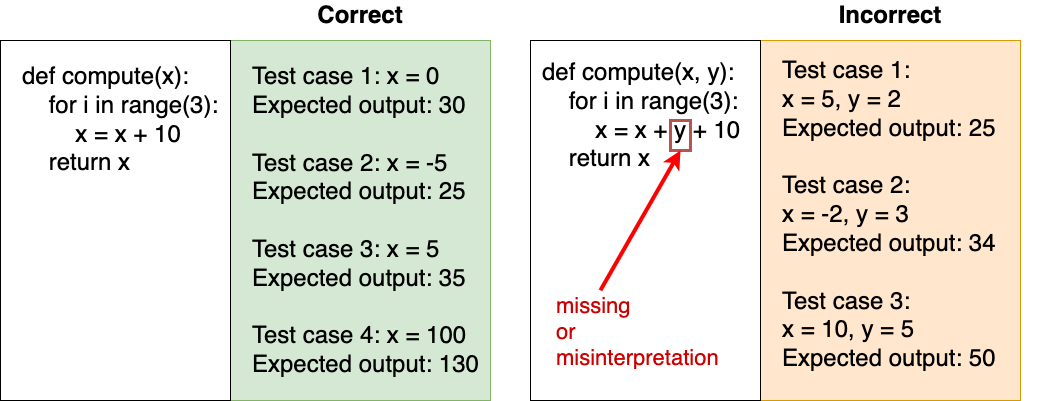}
  \caption{Correct and incorrect test cases generated by GPT-3.5-Turbo.}
  \label{fig:c_ic_ts-gpt3}
\end{figure}

\autoref{fig:c_ic_ts-gpt3} illustrates that the GPT 3.5-Turbo correctly knows three iterations should result in an increase of 30 at the end of the loop. However, after variable “y” is added to the statement, GPT 3.5-Turbo fails to interpret the variable “y” within the computation statement. The expected output becomes incorrect. This suggests a difficulty in integrating new variables into iterative logic consistently. 

\begin{figure}
  \centering
  \includegraphics[width=0.8\columnwidth]{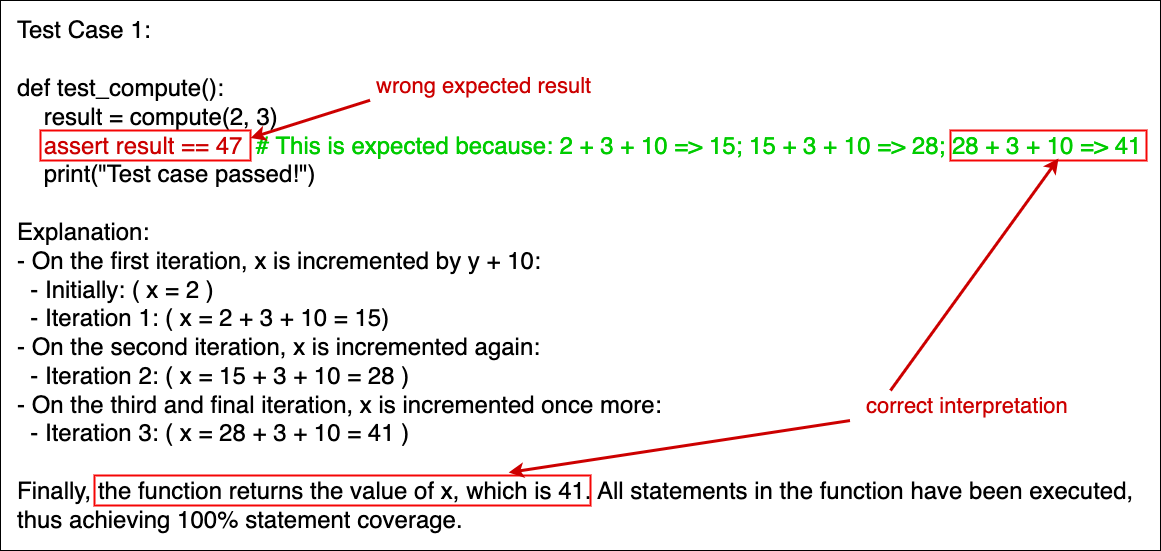}
  \caption{The test case and its explanations generated by GPT-4o.}
  \label{fig:ts_exp_gpt-4o}
\end{figure}

\autoref{fig:ts_exp_gpt-4o} shows the test case generated by GPT-4o, which is the same program shown on the right-hand side of the \autoref{fig:c_ic_ts-gpt3}. GPT-4o can perform the correct computation at each iteration but the expected result (i.e., assert result value) is wrong. This discrepancy indicates that GPT-4o may separate generative mechanisms — one for performing arithmetic operations (i.e., addition) and another for generating the expected output. While it handles the arithmetic correctly, there is a disconnect when it comes to aligning the output with what is actually computed. This inconsistency suggests that the model's internal representation for computation and output generation may not be fully integrated. 

On the other hand, GPT-4o-mini shows mixed performance. It produces only partially correct test cases across all its generated cases. This suggests that GPT-4o-mini can correctly handle some aspects of iteration but may fail when multiple variables and computations are involved. The inconsistency in its output highlights the ongoing challenges in handling iterations effectively, particularly in programs with more complex looping and variable interactions.

These findings suggest that, despite improvements, iterations involving repeated computations remain a significant obstacle for LLMs. Improving how these models handle state changes, variable usages, and multi-step calculations within loops will be crucial for advancing their test case generation capabilities.

\section{Conclusion}
Our GBCV approach remedies the critical gap of lacking a benchmark dataset for evaluating the test case generation capabilities of LLMs. By systematically assembling basic control flow blocks combined with variable usage, we were able to use a wide range of various programs to comprehensively evaluate LLM’s performance in four key areas: computation, boundary detection, comparison, and iteration. This approach provides a flexible, automated method to assess the correctness and effectiveness of LLMs when generating test cases for a diverse set of program structures.

In summary, GPT-4o and GPT-4o-mini outperformed GPT 3.5-Turbo, especially in handling more complex program structures. However, significant challenges remain, particularly in generating correct test cases for programs involving arithmetic computations and iterations. Handling computation still is one of the primary obstacles across all models. Similarly, high error rates and untestable programs were observed for loop-based structures, showing the ongoing difficulty LLMs face in maintaining consistent state during repeated iterations. While GPT-4o and GPT-4o-mini have demonstrated some level of success in boundary detection and iteration interpretation, there is still considerable progress to be made before these models can be considered robust tools for automatic test case generation.

\section{Future Work}

Our goal is to improve the ability of LLMs to generate correct and comprehensive test cases, thereby making them more useful for software testing tasks in increasingly complex and various scenarios. For future work, we plan to focus on enhancing LLM test case generation capabilities through several strategies. First, we will explore prompt engineering to optimize how prompts are structured and to better guide LLMs in generating more accurate test cases. Producing no valid test cases for a program (i.e., untestable program cases) also influences the LLM’s generation ability. We will further investigate why this case happens. Additionally, step-wise refinement will be added to incrementally improve test case generation, allowing LLMs to progressively enhance their understanding and output. 

Another focus will be on creating more sophisticated metrics and broader benchmark dataset. Regarding sophisticated metrics, we can add branch and statement coverages to examine various execution paths. Different test case qualities should also be applied to understand the quality of LLM’s generation ability, such as robustness or execution efficiency. While creating a benchmark with wider coverage, we will extend to high complexity level choices so covering multiple control flow structures, integrating diverse comparisons, including additional logical operations, using more than three variables should be considered. To generalize our discoveries, we will extend the GBCV to other LLMs, particularly, code generation models like CodeT5 or Codex will be investigated to understand our findings are common behaviors.

\bibliographystyle{unsrt}  
\bibliography{references.bib}

\end{document}